\documentclass[12pt,preprint]{aastex}
\usepackage{emulateapj5} 
\usepackage{psfig}
\usepackage{amsmath}
\usepackage{amssymb}
\usepackage{amsbsy}
\usepackage{eucal}
\usepackage{amsthm}

\shorttitle{The GraS Hypothesis: Dynamical Analysis in the Small Velocity Regime}
\shortauthors{Marinoni \& Piazza}

\begin{document}

\title{The Gravitational Suppression Hypothesis:\\
Dynamical Analysis in the Small Velocity Regime}

\author{Christian Marinoni$^{1}$  and Federico Piazza$^{2,\,3}$}
\affil{$^{1}$ Laboratoire d'Astrophysique de Marseille,  Traverse du Siphon B.P.8, 13376 Marseille, France\\
         $^{2}$ Dipartimento di Fisica, Universit\`a di Milano Bicocca, Piazza delle Scienze 3, I-20126 Milano, Italia\\
$^{3}$ Institute of Cosmology and Gravitation, University of Portsmouth, Portsmouth PO1 2EG, UK}
\email{Christian.Marinoni@oamp.fr, Federico.Piazza@mib.infn.it}
 
\begin{abstract}
In a previous paper we have proposed a non--Newtonian, phenomenological  
description of the effective gravitational force acting between exotic 
dark matter and baryons, which was shown to fit well the kinematics in 
the inner regions of low surface brightness, gas--rich galaxies. The Gravitational 
Suppression (GraS) scheme is, admittedly, an {\it ad-hoc} model; it does 
however successfully address interesting cosmological issues. Here, we
test the potential generality of the GraS concept. GraS predictions on motions on scales measurable 
in dwarf spheroidal galaxies and in the solar neighborhood show significant departures
from those of Newtonian mechanics. By analyzing such systems, we find
that the paradigm of a universal GraS potential cannot be rejected.
Indeed, compared to Newtonian predictions, GraS provides a better 
description of data, when realistic dark matter density profiles are 
considered.
\end{abstract}

\keywords{Cosmology: dark matter --- Galaxies: dwarf --- Galaxies: halos --- Galaxy: solar neighborhood --- Gravitation --- Stellar dynamics}

\vspace{0.08cm}

\section{Introduction}

Dark matter (DM) has proved to be an essential component 
of the self--consistent cosmological framework, within which 
observations on a wide range of cosmic scales are generally 
interpreted. Despite its pervasive role, however,  its nature 
is still elusive and its properties mostly speculative. In an
attempt to gain insight of its fundamental characteristics,  and
motivated by the apparent conflicts faced by Cold Dark Matter 
(CDM) scenarios when compared with observations on small
cosmological scales \citep[e.g.,][]{SK, OS}, we have recently 
hypothesized that the gravitational attraction between DM and baryons 
may be suppressed on the kpc scale \citep[][hereafter paper I]{PM}. 
The gravitational suppression (GraS) hypothesis was suggested by 
the simple observation that all the various inconsistencies between 
theory and observations appear to be characterized by a common 
physical length. GraS minimally modifies the existing paradigm, 
i.e. it does not alter large--scale interactions or baryon and DM 
self--dynamics. 

In paper I we modeled the  effective force acting between visible 
and dark particles by adding a short--range Yukawa contribution to 
the standard Newtonian potential. We also investigated the consistency 
and universality of such an empirical gravitational paradigm, fixing 
the model parameters using observations of the rotation curves of low 
surface brightness (LSB), gas--rich galaxies, the cores of which are 
thought to be DM--dominated. The scale length $\lambda$ over which the 
Yukawa contribution is effective is about one kpc, and its strength on 
small distances is equal and opposite to the Newtonian one. In other 
words, the inner dynamics of LSB galaxies seems to point towards a 
``total suppression'' scenario, whereby DM is completely oblivious of 
normal matter in its immediate vicinity.

The analysis of paper I was mainly concerned with rotationally
supported  disk galaxies having circular velocity of order $\sim 100$ km/s. 
Here, we investigate consequences and  predictions of the proposed gravity  
model on a class of systems having different matter distributions and 
dynamical symmetries, with characteristic velocities nearly a factor 10 
smaller than those of typical LSB galaxies. In particular we aim:

a) to assess the ability of the model in describing the kinematics of 
    dwarf spheroidal (dSph) galaxies, i.e. in a physical regime of 
    spherically symmetric,  pressure-supported, low-velocity, virialized 
    systems. This study is motivated by recent claims \citep[e.g.,][]{lok}
    according to which a universal Navarro, Frenk \& White (1997, NFW) 
    DM profile can reproduce the shape of the velocity dispersion profiles 
    of dSph galaxies only if a substantial (and poorly justified) amount 
    of tangential anisotropy in the velocity distribution of dSph stars is
    assumed. 
    By confronting kinematical data of dSph galaxies with the GraS 
    predictions, we wish to verify whether the latter achieves the
    same interpretative success as with LSB galaxies

b) to check the ability of GraS in describing both kinematics and space 
    distribution of stars  perpendicular to the galactic disk in the vicinity 
    of the Sun \citep[e.g.,][]{Oo, SBS}. Since the presence of a spheroidal,  
    non baryonic, DM halo embedding the thin galactic disk 
    affects the density distribution of stars near the galactic plane \citep{bah}, 
    it is imperative to quantify the dynamical effects on stellar motions  
    which are the consequence of GraS. It has been noted \citep{kui} that 
    an intrinsically different non-Newtonian theory (Modified Newtonian Dynamic: MOND, by Milgrom 1983), 
    while capable of explaining the flat rotation curves of galaxies without 
    requiring the presence of a dark matter halo, leads to inconsistent predictions 
    on the vertical component of the gravitational acceleration experienced 
    by solar neighborhood disk stars. Does GraS score any better?

\section{The Model}

As in paper I, we model the proposed short-range modification of gravity 
by adding to the standard Newtonian potential a repulsive Yukawa contribution which 
is active only in the mixed (dark--visible) sector. We can then write the 
total gravitational potential acting on baryons as
$\Phi =  \Phi_N - \Phi_Y$, where $\Phi_N$ satisfies the usual Poisson equation 
\begin{equation}\label{1}
\nabla^2 \, \Phi_N = 4 \pi G \, (\rho_B + \rho_D)
\end{equation}
where $\rho_B$ and $\rho_D$ are respectively the baryonic and DM densities. 
The Yukawa contribution $\Phi_Y $ typically dies out exponentially with a 
radial scale length $\lambda$, according to   
\begin{equation}\label{22}
(\nabla^2 - \lambda^{-2})\, \Phi_Y = 4 \pi G \, \rho_D.
\end{equation}
Notice that  $\Phi_Y$ is sourced by the DM energy density only. The resulting 
gravitational potential between a DM and a baryonic particle at a distance $r$ 
from each other goes like $(e^{-r/\lambda} -1)/r$. In what follows we assume 
$\lambda = 1$ kpc as suggested by the study of LSB galaxies in paper I.

\section{Mass distribution and Dynamics in dSph Galaxies}

Dwarf spheroidal galaxies, being small and DM dominated, provide a test for 
the validity of GraS in a low-velocity, shallow potential regime.
Since the typical luminous dimension of a dSph is $\sim 1$ kpc,  its baryonic 
dynamics is expected to be strongly influenced by the specific form  of our 
gravitational potential,  which is significantly dampened with respect 
to the pure Newtonian one on these scales. Thus: given a realistic DM density 
profile, is the baryon dynamics derived within the GraS paradigm in agreement 
with the observations of the stellar velocity dispersion in dSph galaxies?

The stellar radial velocity dispersion profile $\sigma_r$ induced  by the total 
potential $\Phi$  can be obtained by solving the  Jeans equation for a pressure 
supported spherical system 
\begin{equation}
\frac{d}{dr}(\rho_B \sigma_r^2) + 2 \frac{\beta}{r} \rho_B \sigma_r^2+\rho_B \frac{d \Phi}{dr}~=~0
\label{3}
\end{equation}
where $\beta=1-\sigma_{\theta}^2(r)/\sigma_{r}^2(r)$ accounts for the
anisotropy in the stellar velocity field.

The luminous component of a dSph galaxy  is well described by the 
S\'ersic profile  \citep{ser, cio}  
\begin{equation}
I(r)~=~I_0 \exp[-(r/r_0)^{1/m}]
\end{equation} 
with exponential index $\sim 1$ (but see Caldwell 1999) where $I(r)$ is the 
surface brightness. The baryonic energy density $\rho_B(r)$ can be  obtained 
from $I(r)$  by deprojection:
\begin{equation} 
\rho_B(r)=-\frac{\Upsilon_{*}}{\pi} \int_{r}^{\infty} \frac{dI(t)}{dt}\frac{dt}{\sqrt{t^2-r^2}}
\label{6}
\end{equation}
(see \citet{maz} for an explicit analytical solution of this integral) 
where $\Upsilon_{*}$ is the baryonic mass-to-light ratio. To first approximation, 
$\Upsilon_{*}$ is independent on the radius for the objects considered. The 
solution \eqref{6} can be applied to equation \eqref{3} and in the inversion
of equation \eqref{1} to obtain $\Phi_N$. The DM component $\rho_D$ is also
needed, as it  contributes to both the Newtonian \eqref{1} and the Yukawa 
\eqref{22} potentials. We describe $\rho_D$ via a generalized NFW profile 
of inner slope index $\gamma$ 
\begin{equation}
\rho_D(r)=\frac{\rho_0 \, \delta}{(r/r_s)^{\gamma} (1+r/r_s)^{3-\gamma}},
\label{7}
\end{equation}
where $\rho_0$ is the critical density of the Universe, $\delta$ is a 
characteristic overdensity and $r_s$ a radial scale parameter. According 
to N-body simulations, $\gamma =1$ \citep{NFW} and $\gamma = 1.5$ \citep{MOO2} 
bracket the plausible range of variation of this parameter \citep{nav}.

The virial radius $r_v$ of an object is that within which the  mean density is 
$\Delta \simeq 200 \rho_\circ \simeq 0.023 (M/h^{-1} M_\odot)^{1/3}$ kpc. 
The radius $r_s$ marks the transition from the outer $r^{-3}$ to the inner 
$r^{-\gamma}$ density behavior. A concentration parameter for the DM halo
$c=r_v/r_s$ can be defined, which in N--body simulations exhibits  
a mild dependence on the total mass of the halo. We have not found explicit
results for objects of masses $< 10^{11} M_\odot$, so we extrapolate the results 
of \cite{ENS} to small masses and use, for the NFW $\gamma = 1$ case, the formula
\begin{equation} \label{scaling2}
c_{\gamma =1} = 81\, \left(\frac{M}{ h ^{-1} M_\odot}\right)^{-0.07} .
\end{equation}
while, for the steeper profile $\gamma = 1.5$ we set  
$c_{\gamma = 1.5} =  (1/2)c_{\gamma =1}$ \citep{JS}.

Equation \eqref{22} in the case of spherical symmetry reduces to 
\begin{equation} \label{spherical}
\frac{d^2 \Phi_Y}{d r^2} + \frac{2}{r} \frac{d \Phi_Y}{d r} - \frac{\Phi_Y}{\lambda^2} = 
4\pi G \rho_{D} (r).
\end{equation}
and in paper I an integral expression for $\Phi_Y$ in terms of $\rho_{D}$ 
was outlined. However, for generic $\gamma \neq 1$ only numerical solutions 
of \eqref{22} can be found, while the solution of the Poisson equation \eqref{1} 
can always be given in closed analytic form for any value of $\gamma$. Once the 
total potential $\Phi$ is obtained, equation \eqref{3} can be solved for the 
radial velocity dispersion. A more useful quantity is the line-of sight velocity 
dispersion $\sigma_{los}$, which is related to $\sigma_{r}$ through \citep{bin}
\begin{equation}
\sigma^{2}_{los}(r) =\frac{2}{\Upsilon_{*} I(r)}\int_{r}^{\infty} \!\!\left( 1-\beta\frac{r^2}{t^2} \right) 
\frac{\rho_B(t) \sigma_r^2(t) t}{\sqrt{t^2-r^2}}dt
\label{4}
\end{equation}
and provides a direct comparison with observations.

\section{Models vs. Observations}

Spatially resolved kinematic data became recently available for two dSph 
galaxies: Fornax (Mateo 1997) and Draco (Kleyna 2002). Under the assumption 
that these galaxies are in dynamical equilibrium, as found by 
\cite{MAT98, pia, ode, kle}), we can  compare the observed velocity 
dispersions with those predicted by GraS. 
Assuming  that the measurement errors of the dSph  velocity profiles 
are normally distributed (possible systematics such as the influence 
of binary populations are negligible as shown for example by  
\cite{der}), we compare the observations of
$\sigma_{los}$ with Equation (9) and constrain the scale parameters
of the DM density profile via least square minimization. 
We also fix the parameters characterizing the stellar distribution 
(i.e. the baryons properties) as given by the observations, as
shown in Table 1. 

\vbox{
\begin{center}
\leavevmode
\hbox{
\epsfxsize=9.1cm
\epsffile{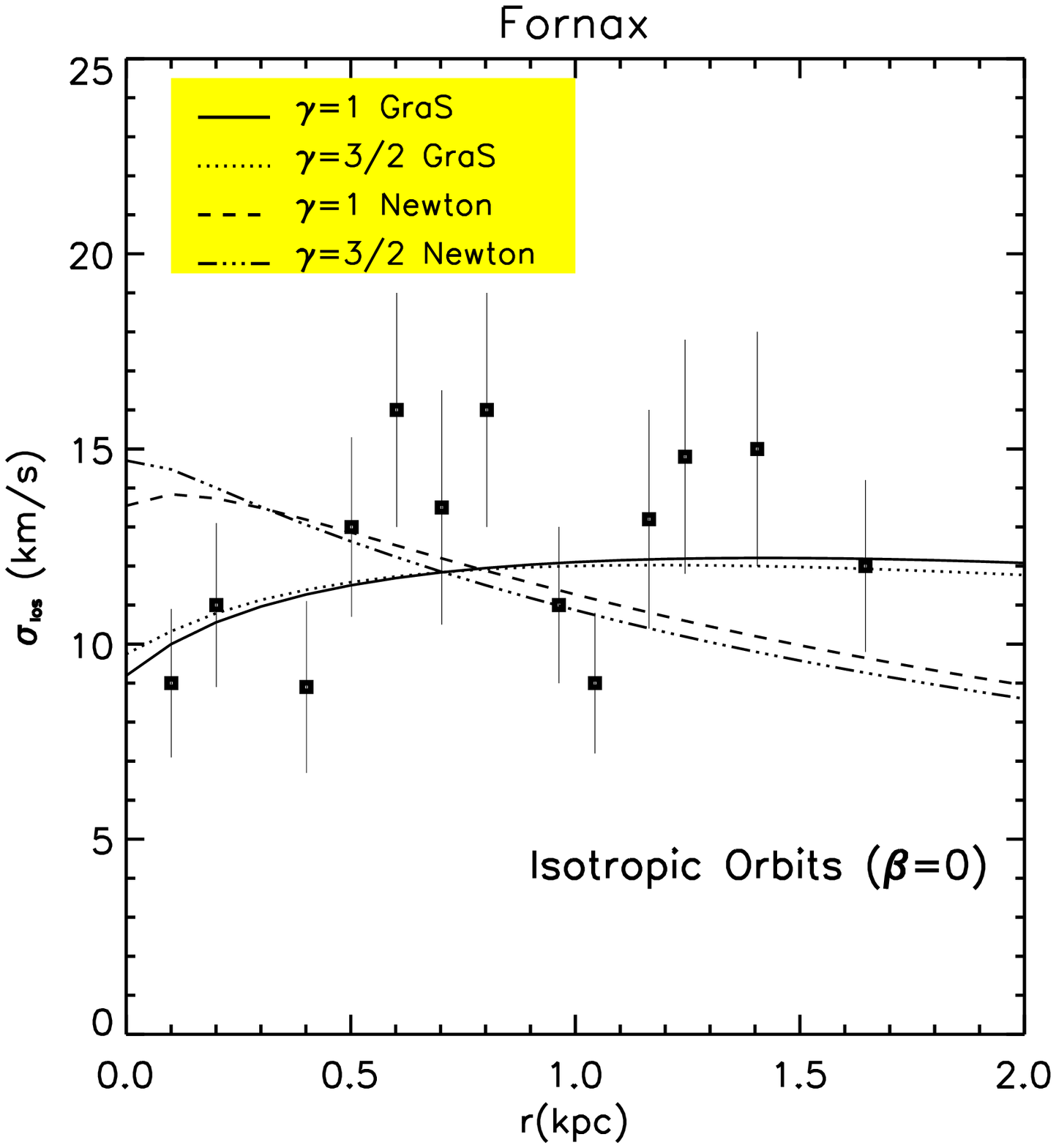}}
\begin{small}
\figcaption{
The observed velocity dispersion profile of Fornax \citep{MAT97} 
is shown  together with four different  best fitting velocity dispersion profiles inferred 
assuming isotropic orbits ($\beta$ is kept fixed to 0 in the fit).
Theoretical predictions are computed  in the case of  a DM inner density profile slope of 1 
\citep{NFW}  or 3/2 \citep{MOO2}  
and according to both the pure Newtonian and the GraS paradigms, as labeled.
Error bars indicate 1$\sigma$ uncertainties.
\label{fig1}}
\end{small}
\end{center}}

Numerical minimization is performed assuming as free parameters of the 
model  the virial mass $M$ of the dark halo embedding the dSph and the 
velocity anisotropy parameter $\beta$  which, for simplicity, is assumed 
to be independent of the radial scale. Thus, our results do not rely on 
any {\it a-priori} assumed anisotropic model: the velocity anisotropy is 
self-consistently evaluated by our fitting scheme.
The resulting best fitting values are presented in Table 2.

Our results may be described as follow: 

a)   the predicted  GraS gravitational  acceleration generated 
by a two component fluid of baryons (distributed according to the Sersic
profile) and DM (distributed according to the generalized NFW profile) 
is in agreement with the stellar velocity dispersions observed in the 
dSph galaxies  Fornax and Draco, if the stellar orbits are isotropic
as shown in  Figs. 1 and 2

b) the best fitting velocity anisotropy parameter $\beta \sim 0$ derived 
within  the GraS gravitational paradigm
is  consistent with the results of N-body simulations of DM halos 
\citep[e.g.,][]{tho}. Table 2 shows  that a Newtonian analysis of the 
kinematics within a cuspy, NFW  DM profile agrees with the observations
only if the existence of a strong tangential anisotropy in the dSph stellar 
orbits is assumed (see also \citet{lok}). The amount of tangential anisotropy needed to 
fit the profiles is, however,  unphysically large if confronted with  
observations of stars in the solar neighborhood \citep[e.g.,][]{chi} or 
the evidence that elliptical galaxies are moderately, radially anisotropic 
\citep[e.g.,][]{ger}
\vbox{
\begin{center}
\leavevmode
\hbox{
\epsfxsize=9.1cm
\epsffile{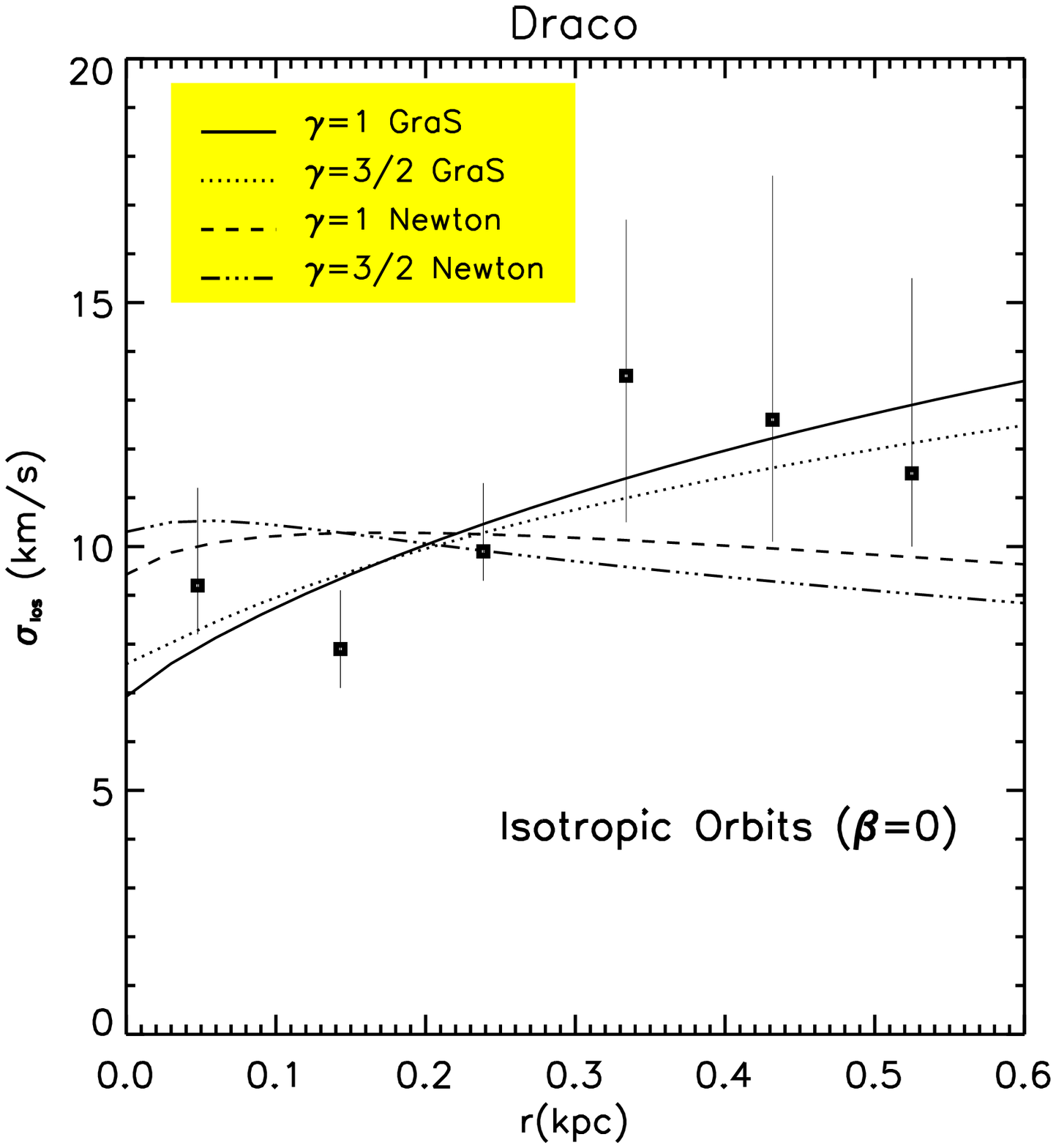}}
\begin{small}
\figcaption{
The same as in fig. 1 but for the Draco  velocity dispersion profile.
Data are from \cite{kle}
\label{fig2}}
\end{small}
\end{center}}

\vspace{0.5cm}

c)  the best fitting mass within the virial radius 
depends on the assumed concentration parameter  (eq. 7).
For realistic values of $c$,  and for all gravity models, the mass of 
Fornax and Draco is  larger than commonly assumed (see  \citet{sto}
for a similar conclusion).
Since in pressure supported systems 
there is a well known degeneracy of density profiles versus velocity 
anisotropy \citep{bin2}, the best fitting  masses present a non negligible  
degree of covariance with respect to the best $\beta$ estimate.
The more negative the $\beta$ parameter (as in the Newtonian case), the 
bigger is the inferred mass (see Table 2). Moreover, the steeper the cusp 
of the profile (higher $\gamma$), the more tangential is the velocity
distribution required to fit the data, in agreement with the finding of 
\cite{lok}. We also note that our results for M$_{t}$/L$_{V}$ (the 
mass-to-light ratio inside the dSph tidal radius $r_t$)  agree in order of 
magnitude with other independent estimates based on Newtonian analysis 
(e.g., Kleyna et al. (2001,2002); \cite{ode})

d) dSph galaxies have low densities and therefore they lie in the regime 
of small accelerations for which MOND provides an alternative interpretation 
of the observed kinematics, without reliance on the DM hypothesis.
However, Lokas (2001) and Kleyna et al. (2001) found that  the best fitting 
values of $a_\circ$,  the characteristic acceleration scale  
(a$_0 \sim 10^{-8}$ cm s$^{-2}$) below which the laws of Newtonian dynamics 
change, are different for different dSph (low for Fornax, much higher than 
expected in the case of Draco). They concluded that the Draco poses a serious 
problem for MOND. At variance, the GraS parameters are stable and robust  
even in the case of these two systems which have similar velocity dispersion
but very different luminosities.

\section{The solar neighborhood dynamics}

We next test GraS predictions in the sub-kiloparsec, small velocity regime
of stellar kinematics in the direction $z$ perpendicular to the galactic 
disk. In the vicinity of the disk the gravitational potential can be 
locally described by \citep{bah}
\begin{equation} \label{arrara}
\frac{\partial^2 \Phi_N}{\partial z^2} =4\pi G (\rho_B^{\rm disk} + \rho_D^{\rm halo}) .
\end{equation} 
In what follows, derivations with respect to $z$ and $r$ are meant while 
keeping the other coordinates -- cylindrical and spherical respectively -- 
fixed. Moreover, equation \eqref{arrara} is derived under the assumption  that
the rotation curve is flat ($\rho_D^{\rm halo} \propto  r^{-2}$) which is, to 
a good approximation, a valid assumption at our position in the Galactic disk. 
We also note that in the solar neighborhood the halo density is about one tenth 
of the disk baryonic density \citep{GGT}. As a consequence, the halo effects on 
the nearby stellar dynamics can be treated as a perturbation, approximated
by expanding in $\epsilon \equiv \rho_D^{\rm halo}/\rho_B^{\rm disk}$ \citep{bah}.
The dynamics of a single isothermal population of density $\rho$ with velocity 
dispersion $\sigma_z$ is then described by the first moment of the Boltzmann 
equation $\sigma_z^2 \partial \rho/\partial z = - \rho \partial \Phi/\partial z $.

We next estimate the correction terms introduced by GraS into eq. \eqref{arrara}. 
Note that, since $\Phi_Y$ is spherically symmetric, 
\begin{equation} \label{expansion}
\frac{\partial^2 \Phi_Y}{\partial z^2}~=~\frac{1}{r} \frac{d \Phi_Y}{d r} \,+\, 
{\cal O}\left(\frac{z}{r}\right)^2 \, \frac{d^2 \Phi_Y}{d r^2},
\end{equation} 
where the height from the disk  $z$ is small with respect to our distance $r$ 
from the center of the Galaxy. In order to estimate the first term on the right
hand side of \eqref{expansion} we use equation \eqref{spherical}. Since the
distance of the Sun from the center of the Galaxy is about 8.5 kpc, we  
expand the solution in the $r \gg \lambda$ limit to obtain
\begin{equation} \label{ararara}
\Phi_Y(r)~=~ - 4\pi G \lambda^2 \rho_D^{\rm halo}(r) \left[1 + {\cal O}\left(\frac{\lambda}{r}\right)^2\right] .
\end{equation}
Deviations from the Newtonian equation \eqref{arrara} are therefore
expected at  order $(\lambda/r)^2$ in the halo density contribution. 
For instance, in the isothermal approximation $\rho_D^{\rm halo} \propto r^{-2}$,
we obtain
\begin{equation}
\frac{\partial^2 \Phi}{\partial z^2} ~\simeq~4\pi G \rho_B^{\rm disk} \left[1 + \epsilon \left(1 - \frac{2\lambda^2}{r^2}\right)\right] .
\end{equation} 
At the Sun's location, the correction to the Bahcall parameter $\epsilon$ is 
thus negligible and of order $\sim 3 \%$. We conclude that GraS is fully 
compatible and  does not compromise our current understanding of the kinematics 
in the solar neighborhood. 

GraS offers a phenomenologically viable paradigm with which to 
reconcile mismatches between observations and CDM predictions on
small cosmological scales.

\acknowledgments
We thank R. Giovanelli for inspiring comments.
We also acknowledge discussions with  S. Colombi, J.M. Deharveng, 
O. Le F\`evre, G. Mamon, A. Mazure and D. Wands.
CM acknowledges financial support from the Centre National de la Recherche Scientifique and the 
Region PACA. FP acknowledges financial support by INFN, MURST and the European
Commission RTN program HPRN-CT-2000-00131 in association with the University of Padova, 
and the Laboratoire d'Astrophysique de Marseille for hospitality.

\clearpage

\begin{deluxetable}{lcc}
\tablewidth{0pc}
\tablecaption{Dwarf parameters}
\tablehead{
\colhead{Parameters}  &
\colhead{Fornax}  &
\colhead{Draco} \\
}
\startdata 
$r_0$(kpc) &  0.35 & 0.18 \\
$r_t$(kpc) &    2.85   &  0.95     \\
m              & 1 & 1.2 \\
distance (kpc) & 138  & 82  \\
$L_{V}(L_{\odot})$  & $ 1.55 \cdot 10^{7}$ & $2.6 \cdot 10^{5}$ \\

$\Upsilon_{*} (M_{\odot}/L_{\odot})$ & 2.5 & 2.5 \\
\enddata
\tablecomments{ All data are taken from \cite{MAT98} except the Sersic scale
radius, the exponential index m of the luminosity density profile and the tidal radius of Draco  which have been 
recently re-estimated by
\cite{ode} using SDSS data. The baryonic mass-to-light radius represents the average
value for a typical globular-cluster star population \citep[e.g.,][]{MAT91,pry}}
\end{deluxetable}

\begin{deluxetable}{llccccc}
\tablewidth{0pc}
\tablecaption{Best Fitting parameters}
\tablehead{
\colhead{ }  &
\colhead{DM inner}  &
\colhead{Gravity Model} &
\colhead{$\beta$} &
\colhead{$M $}  &
\colhead{$M_{t}/L_V$}  &
\colhead{$\chi_{\nu}^2$} \\
\colhead{ }  &
\colhead{slope}  &
\colhead{ } &
\colhead{ } &
\colhead{$(10^{9}M_{\odot})$}  &
\colhead{$10^{2}M_{\odot}/L_{\odot}$ } &
\colhead{  } \\
}
\startdata 
Fornax & $\gamma=1$    & GraS     &$0^{+0.2}_{-0.4}$ & $7.5^{+3}_{-2.5}$  & 0.65 &0.96 \\[1.4mm]
           & $\gamma=3/2$ & GraS     &$-0.1^{+0.2}_{-0.2}$ &$9.5^{+4}_{-3}$   & 0.65 &0.95 \\[1.4mm]
           & $\gamma=1$    & Newton &$-1.5^{+0.9}_{-1.8}$ &$1.1^{+0.4}_{-0.4}$ & 0.23  &  0.91 \\[1.4mm]
           & $\gamma=3/2$ & Newton &$-3^{+0.7}_{-0.8}$&$13.6^{+5}_{-4}$   & 0.76 & 0.92\\[1.4mm]
\hline      \\                                           
Draco  & $\gamma=1$     & GraS     &$0.2^{+0.18}_{-0.23}$&$30^{+25}_{-15}$&16 &0.78\\[1.4mm]
           & $\gamma=3/2$ & GraS     &$0.2^{+0.11}_{-0.16}$&$18^{+7}_{-6}$ &15  &0.80\\[1.4mm]
           & $\gamma=1$     & Newton &$-1.4^{+0.9}_{-2.5}$&$1.7^{+1.1}_{-0.7}$  &5 &0.78\\[1.4mm]
           & $\gamma=3/2$ & Newton &$-6.4^{+1.5}_{-1.5}$&$92^{+80}_{-50}$  & 33  &1.07\\[1.4mm]

\enddata
\tablecomments{All the uncertainties represent 1$\sigma$ errors}
\end{deluxetable}

\end{document}